\documentclass[24pt]{article}
\usepackage{amssymb}
\usepackage{epsfig,caption,graphicx,eepic,epic}
\usepackage{pstricks}
\usepackage{amssymb,amsmath}
\usepackage{multido}
\usepackage{array}
\begin{document}
\def\be{\begin{equation}}
\def\ee{\end{equation}}
\def\ba{\begin{eqnarray}} 
\def\ea{\end{eqnarray}}
\def\nn{\nonumber}
\newcommand{\bbf}{\mathbf}   
\newcommand{\rrm}{\mathrm}
\title{\bf Evolution of open quantum systems: time scales, stochastic and continuous processes}
\author{Tarek Khalil$^{a}$
\footnote{E-mail address: tkhalil@ul.edu.lb}\\ 
and\\
Jean Richert$^{b}$
\footnote{E-mail address: j.mc.richert@gmail.com}\\ 
$^{a}$ Department of Physics, Faculty of Sciences(V),\\
Lebanese University, Nabatieh,
Lebanon\\ 
$^{b}$ Institut de Physique, Universit\'e de Strasbourg,\\
3, rue de l'Universit\'e, 67084 Strasbourg Cedex,\\      
France} 
 
\date{\today}
\maketitle 
\begin{abstract}
The study of the physical properties of open quantum systems is at the heart of many present investigations which aim to describe their dynamical evolution, on theoretical ground and through physical realizations. Here we develop a presentation of different aspects which characterize these systems  and confront different physical situations which can be realized leading to systems which experience Markovian, non-Markovian, divisible or non-divisible interactions with the environments to which they are dynamically coupled. We aim to show how different approaches describe the evolution of quantum systems subject to different types of interactions with their environments.
\end{abstract}


\section{Introduction}
                       
Quantum systems are generally in contact with physical environments which may be of different types. The understanding and description of these sytems have been the object of a  multitude of studies, see f.i. the recent report ref.~\cite{riv1}. The existence of an environment induces exchange processes such as energy, heat, the measurement of physical observables  which characterize the system. These processes are characterized by different types of properties. They are either induced by discontinuous stochastic or deterministic continuous interactions. In general the interactions induce a time delay between the environment and the system, optimally they may be close to instantaneous. The understanding and control of these processes is of paramount importance for the realization 
of quantum objects and the measurement of their physical properties in many fields of quantum technology,
see f.i. ~\cite{rai,leib,blo,fu,lad}.   

In the present work we aim to present the different physical cases concerning the dynamical evolution of open quantum systems, confront different approaches some of which have already been examined in this field. We try to show the consequences of the nature of the interaction which couples them to their environments.\\  
                     
The content of the present work is the following. In section 2 we recall the essential mathematical definitions of a Markov process. In section 3 we show by means of a model physical and temporal conditions which must be realized in order to generate a physical Markovian quantum system. Section 4 deals with systems which are explicitly subject to stochastic interactions with their environment  and show the conditions under which divisibility is realized. In section 5 we develop the central argument that divisibility may also be reached when the environment and the interaction beween the environment and the system is deterministic. We work out the conditions under which condition this is realized. Section 6 is devoted to a summary of the results and some further comments. Explicit calculations are developed in the appendices.                       
                       
\section{The interaction of the system with its environment is of stochastic nature}
                       
\subsection{Classical Markov processes - Mathematical definition}                       

We recall here some mathematical aspects of Markov processes and their relationship with the so called physical Markovian property in order to show the link with its relationship with its use for the description of physical systems. The original paper concerning this concept by A. A. Markov was published in russian in the Bulletin of the Mathematical and Physical Society of the University of Kazan, 1906. An english translation can be found in ~\cite{mark}. The mathematical concepts developed in the present section have been taken from ref.~\cite{oli}.\\  
 
\subsection {Random variables}

Consider a sample space $\Omega$ of possible outcomes of a random process ${w_{i}}$. Each outcome is an event. The assignation of a real number to each $w$ leads to a random variable $X(w)$, a single-valued real function of $w$ and $\Omega$ is the domain of $X$.

\subsection {Probabilities}

Consider a random variable $X$ and $x$ a fixed real number, $A_{X}$ the subset of $\Omega$ which consists of all real sample points to which $X$ assigns the number $x$ 

\ba
A_{X}=[w | X(w)=x]=[X=x]
\label{eq1}
\ea

Because $A_{x}$ is an event it will have a probability $p=P(A_{x})$. One defines a cumulative distribution function as
  
\ba
F_{X}(x)=P[X \leq x] 
\label{eq2}
\ea
 with $x$ in the interval $[- \infty, + \infty]$.  
  
\subsection {Stochastic or random process}

Consider a set of random variables depending on a continuous variable $t$. Define $X(t,w)$ as a collection of time functions for a fixed value of $w$.

A stochastic or random process is a family of random variables $[X(t,w)]$ defined over a given parameter set $T$ indexed by $t$.  

In the following the fixed event parameter $w$ will be left out in the notations.

\subsection {Markov process and strong Markov process} 

\begin{itemize}

\item A stochastic process $[X(t)]$ where $t$ belongs to a continuous ensemble $T$ is called a 1st order Markov process if for a sequence $[t_{0}, t_{1},...,t_{n}]$ the conditional cumulative distribution function $F_{X}$ of $X(t_{n})$ for a given sequence $X(t_{0})$, $X(t_{1})$,..., $X(t_{n-1})$ depends only on $[X(t_{n-1})]$. The conditional probability $P$ for generating a random value $X(t_{n})$ at time $t_{n}$ if its value was  
$X(t_{n-1}),.....,X(t_{1}),X(t_{0})$ at times $t_{n} \geq t_{n-1} ,...,\geq t_{1} \geq t_{0}$ is given by

\ba
P[X(t_{n})| X(t_{n-1}),.....,X(t_{1}),X(t_{0})]= P[X(t_{n})| X(t_{n-1})]
\label{eq3}
\ea  
  
\item The process is strong if $[X(t+s)-X(t), s \geq 0]$  has the same distribution as the process
$[X(s),s \geq 0]$ and is independent of the process $[X(s),0 \leq s \leq t]$ i.e., if the process is known at time $t$ the probability law of the future change of state of the process will be determined as if the process started at time $t$, independently of the history of the process between $t=0$ and $t$.\\

\end{itemize}

One can interpret the Markovian property of a system $S$ which evolves stochastically as a loss of memory of the system over an arbitrarily short interval of time. Hence the evolution is governed by a process such that this evolution at any arbitrary time $t$ depends only on $t$, independently of its evolution in the past. The strong limit of the process shows that the evolution is invariant under time translation, it depends only on the time interval between two random events and not of the initial time at which the process is observed.\\

The consequence of the specific time dependence of the Markov assumption applied to physical systems is the divisibility (semi-group) property which will be shown in the following section.  

\section{Physical stochastic quantum processes - Phenomenological approach}

We use here a phenomenological approach ~\cite{sh} which exemplifies the conditions under which a physical system coupled to its environment evolves in such a way that the characteristic time  of its evolution is independent from memory effects that may be induced by its environment.   

\subsection{Conditions for the existence of a Markovian master equation}

Consider an open quantum system $S$ coupled to its environment $E$. In general $S$ is described by its time-dependent density operator $\hat \rho_{S}(t)$ applied to its environment and can be obtained as the solution of a master equation. The evolution can be followed in terms of the differential equation

\ba
\frac{d\hat \rho_{S}(t)}{dt}=lim_{\tau \rightarrow  0} \frac{\hat \rho_{S}(t+\tau)-\hat \rho_{S}(t)}{\tau}
\label{eq4}
\ea
where $\hat \rho_{S}(t+\tau)=\hat L_{t,\tau} \hat \rho_{S}(t)$, $\hat L_{t,\tau}$ being the evolution operator from $t$ to $t+\tau$. 

Since $S$ is coupled to $E$ the total system $S+E$ is described by a density operator $\hat \rho_{SE}(t)$ whose general expression can be written   

\ba
\hat \rho_{SE}(t)=\hat \rho_{S}(t)\hat \rho_{E}(t)+\delta\hat \rho_{SE}(t)
\label{eq5}
\ea
$\hat \rho_{S}(t)=Tr_{E}[\hat \rho_{SE}(t)]$ and $\hat \rho_{E}(t)=Tr_{S}[\hat \rho_{SE}(t)]$.
The traceless operator $\delta\hat \rho_{SE}(t)$ is generated by the coupling between $S$ and $E$. It perturbs the free evolution of $S$ and may induce retardation effects in the process, hence introduce a memory effect into the description of the evolution. Eq.(4) will show Markovian properties if two conditions are satisfied:
 
\begin{itemize}

\item (a) $\delta\hat \rho_{SE}(t)/\hat \rho_{SE}(t) \ll 1$ 

\item (b) $\| \hat \rho_{E}(t) \| \simeq \bar \rho_{E} $ 

\end{itemize}
where $\bar \rho_{E}$ is a constant density. Relation (a) originates from the fact that the correlations induced by the coupling are considered to be weak, (b) expresses the fact that $E$ is stationary and as a consequence, 
$L_{t,\tau}$ must be independent of $t$ so that $E$ does not depend on earlier times.

\subsection{Time scales}

The system $S$ is characterized by a typical evolution time $\tau_{S}$. Consider the evolution of the phases of $S+E$ as a succession of phases which accumulate coherently over time intervals $\tau_{E}$, the typical memory time of $E$. The process operates as a random walk over large time intervals $t$ during which the phases $\Phi(t)$ add up quadratically as

\ba
\Delta^{2} \Phi(t \gg \tau_{E}) \sim (|H_{SE}|\tau_{E}/\hbar)^{2} t/\tau_{E}=t/\tau_{S}
\label{eq6}
\ea
where $|H_{SE}|$ is the strength of the interaction coupling between $S$ and $E$ and the time $\tau_{S}=\hbar^{2}/(|H_{SE}|^{2}\tau_{E})$ is the typical time over which the system $S$ evolves. From this expression one can see that the time $\tau_{S}$ is much longer than $\tau_{E}$ when the coupling is weak and 
$\tau_{E}$ small. Then one can rewrite expressions of eq.(6) as

\ba
\| \delta\hat \rho_{SE}(t) \|=O(\tau_{E}/\tau_{S})
\notag\\
\| \hat \rho_{E}(t) \|=\bar \rho_{E}+O(\tau_{E}/\tau_{S})
\label{eq7}
\ea

\subsection{Conditions for the realization of a Markovian process}

The properties mentioned above can be realized under specific conditions. If the environment is large the spectrum has a large extension $\Delta_{E}$ and the density of states generally large too. As a consequence the decay time $\tau_{E}=\hbar/\Delta_{E}$ which is the time over which the correlations generated by the coupling $\hat H_{SE}$ between $S$ and $E$ survive is small. Over this time interval the phase of the wave function changes by an amount of the order of $\tau_{E}|H_{SE}|/\hbar$. As a consequence it comes out that  

\ba
\| \delta\hat \rho_{SE}(t)\|=O(|H_{SE}|^{2}\tau_{E}^{2}/\hbar^{2})
\notag\\
\| \hat \rho_{E}(t)\|=\bar \rho_{E}+O(|H_{SE}|^{2}\tau_{E}^{2}/\hbar^{2})
\label{eq8}
\ea
The inequality $|H_{SE}|\tau_{E}/\hbar \ll 1$ qualifies the Markovian property of the process: the time interval over which the system keeps the memory of its coupling to the environment and the coupling between $S$ and $E$ have to be small, the environment $E$ has to influence $S$ over a very short time interval compared to the characteristic evolution time $\tau_{S}$ of $S$. This leads to the conditions given by Eq.(7).
 
A rigorous Markovian behaviour would correspond to the condition $\tau_{E}=0$.
  
\subsection{Markov processes and divisibility}

The master equation derived above describes a Markovian system because its derivation relies on the short memory behaviour of the correlations which characterize the coupling of the system to its stationary environment. In the limit where $\tau_{E}$ goes to zero the time correlations 

\ba 
C(t,t')=lim_{\tau_{E} \rightarrow  0}\langle \hat H_{SE}(t) \hat H_{SE}(t'=t+\tau_{E})\rangle \propto
\delta(t-t')
\label{eq9}
\ea
hence at time $t'$ the system has lost the memory of its coupling to the environment at time $t$. This effect is general and does not depend on a specific form of the interaction $\hat H_{SE}$. 
 
As a consequence the master equation  which governs the evolution of $\hat \rho_{S}(t)$ depends on a single time variable. The density operator $\hat \rho_{S}(t_{2})$ will be related to 
$\hat \rho_{S}(t_{1})$ for $t_{2} > t_{1}$ by the relation 
$\hat \rho_{S}(t_{2})= \hat \Phi(t_{2},t_{1})\hat \rho_{S}(t_{1})$. Then for any further time interval $[t_{2},t_{3}]$ one will get the property                                    

\ba 
\hat \rho_{S}(t_{3})= \hat \Phi(t_{3},t_{2})\hat \Phi(t_{2},t_{1}) \hat \rho_{S}(t_{1})
\label{eq10}
\ea 
where $\hat \Phi(t',t)$ is the evolution operator of the open system whose properties have been extensively studied
~\cite{gl,hb,vb,ar}.\\

Divisibility is a property of Markov systems, it is governed by time scale considerations. The question which comes next is to know whether this property is specific to these systems, hence if divisibility is equivalent to Markovianity. This will be examined in the recent following approach in which the action of the environment possesses a stochastic character.
 
\section{Formal approach of stochastic quantum processes}

The former developments do not explicitly allude to the concept of stochasticity which is the classical central concept in a Markov process. This concept has been introduced explicitly and developed in recent work by Pollock and coll. ~\cite{pol1,pol2} who worked out a formal presentation of stochastic quantum processes and derived a necessary and sufficient condition which leads to a rigorous description of a Markov process.

Consider a quantum system which evolves from $t=0$ under the action of $r$ possible external devices of different type $d_{j}^{(r)}$ which act at time $t= t_{j}$, f.i. unitary  transformations, interactions with an environment mathematically represented by completely positive maps. 

At each step $j$ one can define $D_{j}=\sum_{r} d_{j}^{(r)}$. If these operations are repeated $k$  times and are uncorrelated one can define a sequence of actions

\ba
{\bf D_{({k-1 \leftarrow 0})}}=\{d_{k-1}^{(r_{k}-1)};.....;d_{1}^{(r_{1})};d_{0}^{(r_{0 })}\}                                    
\label{eq11}
\ea

The evolution of the density operator of the system under the action of the devices will be given by the linear CP (completely positive) map

\ba
\hat \rho_{k}=\hat V_{({k \leftarrow 0})}[{\bf D_{({k-1 \leftarrow 0})}}]                                  
\label{eq12}
\ea
where $\hat V_{({k \leftarrow 0})}$ leads the system from $t=0$ to time $t_{k}$ if the system was given by the correlated or uncorrelated density $\hat \rho_{0}$ at  time $t=0$.

A quantum Markov process may  be characterized by so called "causal breaks": at $t=t_{k}$ the system is reset by means of the external devices and these actions do not depend on the past.

Consider the system at time $t_{l}$: $\hat \rho_{l}= \hat \rho_{l}[{\bf D_{({l-1 \leftarrow 0})}}]$ and the action on the system at some time $k<l$ where a stochastic action of some $d_{k}^{(r)}$ acts with  probability $p_{k}^{(r)}$ corresponding to the positive projection operator $\hat \Pi_{k}^{(r)}$. Then the system is prepared again in a state $P_{k}^{(s)}$ randomly chosen out of a set $[P_{k}^{(s)}]$. The action at time $t_{k}$ breaks the causal link between the past $j<k$ and the future $l>k$. At time $l$ the density operator can be written as
\ba
\tilde \rho_{k}= p_{r} \hat \rho_{l}(P_{k}^{(s)}|\hat \Pi_{k}^{(r)};{\bf D_{(k-1 \leftarrow 0)}})                                                                
\label{eq13}
\ea
which corresponds to the density operator at time $t_{l}$ at the condition that its outcome  at $t_{k}$ is the state $P_{k}^{(s)}$ with probability $p_{r}$ when at  step $k$ it  was $\hat \rho_{k}$. The conditioning argument is fixed by $\hat \Pi_{k}^{(r)}$ and controls ${\bf D_{({k-1 \leftarrow 0})}}$. Forgetting the probability $p_{r}$ which plays no role in the  determination of the Markovian property of the process one gets finally  

\ba
\hat \rho_{l}= \hat \rho_{l}(P_{k}^{(s)}|\hat \Pi_{k}^{(r)};{\bf D_{(k-1 \leftarrow 0)}})                                                                
\label{eq14}
\ea
which is the quantum equivalent of the classical expression.

The process is Markovian if

\ba
\hat \rho_{l}(P_{k}^{(s)}|\hat \Pi_{k}^{(r)};{\bf D_{(k-1 \leftarrow 0)}})= \hat \rho_{l}(P_{k}^{(s)})                                                              
\label{eq15}
\ea
$\forall$ ${ P_{k}^{(s)}}$, $\hat \Pi_{k}^{(r)};{\bf D_{(k-1 \leftarrow 0)}}$ and $\forall$ $(l,k)$. 

From this definition follows a central theorem.

The process is non-Markovian if and only if  there exists at least two different choices of measures (controls) such that after a causal break at time $t_{k}$ 

\ba
\hat \rho_{l}(P_{k}^{(s)}|\hat \Pi_{k}^{(r)};{\bf D_{(k-1 \leftarrow 0)}}) \neq 
\hat \rho_{l}(P_{k}^{(s)}|\hat \Pi_{k}^{'(r')};{\bf D^{'}_{(k-1 \leftarrow 0)}})                                                              
\label{eq16}
\ea
The evolution is Markovian if $\hat \rho_{l}$ is the same for all linearly independent measures (controls).

This result induces two consequences:

\begin{itemize}

\item A given choice of decomposition $D_{j}=\sum_{r} d_{j}^{(r)}$ leads to a classical distribution iff the quantum process is Markovian according to the definition above.

\item A consequence of the Markovian property is the divisibility (1/2-group property) of the process:\\
       if the system  evolves in time as $\hat\rho(t_{l})=\hat T(t_{l},t_{j})\hat \rho(t_{j})$ its evolution obeys also

\ba
\hat\rho(t_{l})=\hat T(t_{l},t_{k}) \hat T(t_{k},t_{j})\hat\rho(t_{j})
\label{eq17}
\ea
for $l>k>j$

\end{itemize}

It is easy to realize that open quantum systems which interact by means of time-independent interactions cannot be Markovian except if the process goes in no  more than two time steps. This means that non-Markovian processes are the general case.

In the phenomenological derivation of a Markov-type master equation we have seen that many constraints have to be introduced in order to approach a Markovian evolution, among them the fact  that the interaction between the system and the environment should get as small as possible. It is also clear that the existence of memory effects is intuitively understandable. The stochastic action of the environment generates perturbations which take some time in order to be absorbed by the system.   

\section{The interaction of the system with its environment is of deterministic nature: conditions for rigorous  divisibility}

Consider the case for which the interaction between $S$ and $E$ does not necessarily follow a stochastic process, hence open quantum systems are not necessarily driven by a Markovian or non-Markovian mechanism. We want to know whether the time evolution of the system possesses the semi-group (divisibility) property which characterizes Markov processes. 

In the sequel we shall show that divisibility can indeed be realized in systems which are not necessarily Markovian.

\subsection{General expression of the density operator in the Liouvillian formalism}
 
An open system $S$ characterized by a density operator  $\hat\rho_{S}(t)$ which evolves in time from $t_{0}$ to $t$ under the action of the evolution operator $\hat T(t,t_{0})$

\ba
\hat\rho_{S}(t)=\hat T(t,t_{0})\hat\rho_{S}(t_{0})
\label{eq18}
\ea

At the initial time $t_{0}$  the system $S$ is supposed to be decoupled from its environment and characterized by the density operator

\ba
\hat\rho_{S}(t_{0})= \sum_{i_{1},i_{2}}c_{i_{1}}c_{i_{2}}^{*}|i_{1}\rangle \langle i_{2}|
\label{eq19}
\ea
and the environment $E$ by

\ba
\hat\rho_{E}(t_{0})= \sum_{\alpha_{1},\alpha_{2}}d_{\alpha_{1},\alpha_{2}}|\alpha_{1}\rangle \langle \alpha_{2}|
\label{eq20}
\ea
where $|i_{1}\rangle, |i_{2}\rangle $ and $|\alpha_{1}\rangle, |\alpha_{2}\rangle$ are orthogonal states in  
$S$ and $E$ spaces respectively, $c_{i_{1}},c_{i_{2}}$ normalized amplitudes and $d_{\alpha_{1},\alpha_{2}}$ weights such that $\hat\rho_{E}^{2}(t_{0})=\hat\rho_{E}(t_{0})$.

At time $t>t_{0}$ the reduced density operator in $S$ space is $\hat\rho_{S}(t)=Tr_{E}[\hat\rho(t)]$ where $\hat\rho(t)$ is the density operator of the total system $S+E$. It can be written as~\cite{vlb}

\ba
\hat\rho_{S}(t)=\sum_{i_{1},i_{2}}c_{i_{1}}c_{i_{2}}^{*}\hat\Phi_{i_{1},i_{2}}(t,t_{0})
\label{eq21}
\ea
with

\ba
\hat\Phi_{i_{1},i_{2}}(t,t_{0})=\sum_{j_{1},j_{2}}C_{(i_{1},i_{2}),(j_{1},j_{2})}(t,t_{0})|j_{1}\rangle_{S}\langle j_{2}|
\label{eq22}
\ea

where the super matrix $C$ reads

\ba
C_{(i_{1},i_{2}),(j_{1},j_{2})}(t,t_{0})=\sum_{\alpha_{1},\alpha_{2},\gamma}d_{\alpha_{1},\alpha_{2}}
U_{(i_{1}j_{1}),(\alpha_{1}\gamma)}(t,t_{0}) U_{(i_{2}j_{2}),(\alpha_{2}\gamma)}^{*}(t,t_{0})
\label{eq23}
\ea

and

\ba
U_{(i_{1}j_{1}),(\alpha_{1}\gamma)}(t,t_{0})=\langle j_{1}\gamma|\hat U(t,t_{0})|i_{1} \alpha_{1}\rangle
\notag\\
U^{*}_{(i_{2}j_{2}),(\alpha_{2}\gamma)}(t,t_{0})=\langle i_{2} \alpha_{2}|\hat U^{*}(t,t_{0})|j_{2} \gamma \rangle
\label{eq24}
\ea

The evolution operator reads $\hat U(t,t_{0})=e^{-i\hat H(t-t_{0})}$ where $\hat H$ is the total Hamiltonian in 
$S+E$ space and the super matrix $C$ obeys the condition 
$\lim_{t\rightarrow t_{0}}C_{(i_{1},i_{2}),(j_{1},j_{2})}(t,t_{0})=\delta_{i_{1},i_{2}} \delta_{j_{1},j_{2}}$.\\

In the present formulation the system is described in terms of pure states. The results which will be derived below remain valid if the initial density operator at the initial time is composed of mixed states 
$\hat\rho_{S}(t_{0})=\sum_{i_{1},i_{2}}c_{i_{1}i_{2}}|i_{1}\rangle _{S}\langle i_{2}|$.\\

One asks now under what conditions the evolution of $\hat\rho_{S}(t)$  will be divisible. 
 
\subsection{Divisibility: the  system is coupled to a unique state of the environment} 

Consider a system which obeys the divisibility criterion ~\cite{vb,ar} 

\ba
\hat\rho_{S}(t,t_{0})=\hat T(t,\tau)\hat T(\tau,t_{0})\hat\rho_{S}(t_{0})
\label{eq25}
\ea
with $\tau$ in the interval $[t_{0},t]$.

The problem is now to find conditions under which the general expression of $\hat\rho_{S}(t,t_{0})$ obeys the divisibility constraint fixed by Eq.(25) at any time $t>t_{0}$.\\ 

For this to be realized the following relation must be verified by the super matrix $C$

\ba
C_{(i_{1},i_{2}),(k_{1},k_{2})}(t,t_{0})= \sum_{j_{1},j_{2}}C_{(i_{1},i_{2}),(j_{1},j_{2})}(t_{s},t_{0})
C_{(j_{1},j_{2}),(k_{1},k_{2})}(t,t_{s})
\label{eq26}
\ea
The explicit form of this equation is worked out in Appendix A. Writing out explicitly the r.h.s. and l.h.s. of  
Eq.(26) in terms of the expression of $C$ given in Eq.(25) for fixed values of $i_{1}$ and $i_{2}$ one finds from inspection of Eqs.(52-53) that a sufficient condition for this to be realized is obtained if there is a unique occupied state $|\eta\rangle$ in $E$ with $d_{\eta,\eta}=1$. This is in agreement with ref.~\cite{sti}.  

\subsection{Generalization to several states in the environment}

It is our aim here to show that the semi-group (divisibility) ptoperty  can be realized even if there is more than one state in $E$ space. To see this we introduce the explicit expression of the master equation which governs an open quantum system in a time local regime.  Its expression reads~\cite{gl,gk,lain,hall,pearl}
\ba
\frac{d}{dt}\hat\rho_{S}(t)=\sum_{n}\hat L_{n}\hat\rho_{S}(t)\hat R_{n}^{+}
\label{eq27}
\ea
where $\hat L_{n}$ and $\hat R_{n}$ are time independent operators.\\

Using the general form of the density operator $\hat\rho_{S}(t)$ given by Eqs. (19-24) 

\ba
\hat\rho_{S}^{j_{1}j_{2}}(t)=\sum_{i_{1}i_{2}}c_{i_{1}}c^{*}_{i_{2}}\sum_{\alpha\alpha_{2}, \gamma}d_{\alpha_{1},\alpha_{2}}\langle j_{1}\gamma|\hat U(t,t_{0})|i_{1} \alpha_{1}\rangle
\notag\\
\langle i_{2} \alpha_{2}|\hat U^{*}(t,t_{0})|j_{2} \gamma \rangle
\label{eq28}
\ea
and taking its time derivative leads to two contributions to the matrix elements of the operator 

\ba
\frac{d}{dt}\rho_{S1}^{j_{1}j_{2}}(t)=(-i)\sum_{i_{1}i_{2}}c_{i_{1}}c^{*}_{i_{2}}\sum_{\alpha_{1},\alpha_{2}}d_{\alpha_{1},\alpha_{2}}
\sum_{\beta \gamma k_{1}}\langle j_{1}\gamma|\hat H|k_{1} \beta \rangle
\notag\\
\langle k_{1}\beta|e^{-i\hat H(t-t_{0})}|i_{1} \alpha_{1} \rangle \langle i_{2}\alpha_{2}|e^{i\hat H(t-t_{0})}|j_{2} \gamma \rangle 
\notag\\
\frac{d}{dt}\rho_{S2}^{j_{1}j_{2}}(t)=(+i)\sum_{i_{1}i_{2}}c_{i_{1}}c^{*}_{i_{2}}\sum_{\alpha_{1},\alpha_{2}}d_{\alpha_{1},\alpha_{2}}
\sum_{\beta \gamma k_{2}}\langle j_{1}\gamma|e^{-i\hat H(t-t_{0})}|i_{1} \alpha_{1} \rangle  
\notag\\
\langle i_{2}\alpha_{2}|e^{i\hat H(t-t_{0})}|k_{2} \beta \rangle \langle k_{2}\beta|\hat H|j_{2} \gamma \rangle 
\label{eq29}
\ea
and                  

\ba
\frac{d}{dt}\hat\rho_{S}^{j_{1}j_{2}}(t)=\frac{d}{dt}[\rho_{S1}^{j_{1}j_{2}}(t) +\rho_{S2}^{j_{1}j_{2}}(t)]
\label{eq30}
\ea
From the explicit expression of the density operator matrix element given by Eqs. (28-30) one sees that the structure of the master equation given by Eq.(27) which induces the divisibility can only be realized if $|\beta \rangle=|\gamma \rangle$. Three solutions of special interest can be found:

\begin{itemize}

\item There is only one state $|\gamma \rangle$ in $E$ space. This result has already been seen on the expression of the density operator above.

\item The density operator $\hat\rho_{S}(0)$ is diagonal in $S$ space with equal amplitudes of the states 
and the states in $E$ space are equally weighed, $\hat \rho_{E}=\sum_{\alpha}d_{\alpha,\alpha}|\alpha\rangle\langle\alpha|$, $d_{\alpha,\alpha}=1/N$ where $N$ is the number of states in $E$ space. See proof in Appendix B. These states called maximally coherent states have been introduced in a study of quantum coherence ~\cite{bau}.

\item  If the environment stays in a fixed state $|\gamma \rangle$, i.e. if the Hamiltonian $\tilde H=\hat H_{E}+\hat H_{SE}$ is diagonal in a basis of states in which $\hat H_{E}$ is diagonal. Then, if the system starts in a given state $|\gamma\rangle$ it will stay in this state over the whole interval of time and the density operator will be characterized by a definite index $\gamma$  
$\hat\rho_{S \gamma}(t,t_{0})$. For an explicit expression of the matrix elements of $\hat \rho_{S}$ see Appendix C. The central  point to notice here is the fact that this happens if $[\hat H_{E},\hat  H_{SE}]$=0.

\end{itemize}

Each of these conditions is sufficient to insure the structure of the  r.h.s. of Eq.(30). The last one is the most general one. The commutation between $\hat H_{E}$ and $\hat H_{SE}$ is a sufficient condition to induce divisibility. However, as claimed in ~\cite{pol3} and shown below, divisibility does not necessarily induce a Markovian behaviour. We shall give a counter example below. 

\subsection{Memory effects and absence of divisibility: two-time approach}

We use now the projection formalism ~\cite{nak,zwa,ck1,veg} and the expression developed in section 5.1 in order to analyze the time evolution of the density operator of the total system $S+E$ 

\ba
\hat \rho(t,t_{0})=\sum_{i_{1},i_{2}}c_{i_{1}}c^{*}_{i_{2}}\sum_{\alpha}d_{\alpha \alpha}U(t,t_{0})
|i_{1} \alpha \rangle \langle i_{2} \alpha| U^{+}(t,t_{0})
\label{eq31}
\ea
We write the expression of $\hat \rho(t,t_{0})$ in a basis of states in which $\hat H_{E}$ is diagonal.

We introduce projection operators $\hat P$ and $\hat Q$ in $E$ space such that

\ba
\hat P \hat \rho(t,t_{0})=\sum_{k=1}^{n}|\gamma_{k}\rangle \langle \gamma_{k}|\hat \rho(t,t_{0})
\notag\\
\hat Q \hat \rho(t,t_{0})=\sum_{l=n+1}^{N}|\gamma_{l}\rangle \langle \gamma_{l}|\hat \rho(t,t_{0})
\label{eq32}
\ea
where $N$ is the total finite or infinite number of states in $E$ space and $\hat P+\hat Q=\hat I$ 
where $\hat I$ is the identity operator.

The evolution of the density operator is given the Liouvillian equation
\ba
\frac { d \hat \rho(t,t_{0})}{ dt}=\hat L(t) \hat \rho(t,t_{0})=-i[\hat H,\hat \rho(t,t_{0})]
\label{eq33}
\ea

Projecting this equation respectively on $\hat P$ and $\hat Q$ subspaces leads to a set of two coupled  equation

\ba
\frac{d \hat P \hat \rho(t,t_{0})}{dt}= \hat P \hat L(t)\hat P \hat \rho(t,t_{0})+
\hat P \hat L(t)\hat Q \hat \rho(t,t_{0}) (a)
\notag\\
\frac{d \hat Q \hat \rho(t,t_{0})}{dt}= \hat Q \hat L(t)\hat Q \hat \rho(t,t_{0})+
\hat Q \hat L(t)\hat P \hat \rho(t,t_{0}) (b)
\label{eq34}
\ea

Choosing $t_{0}=0$ in order to simplify the equations and solving formally the second equation gives
\ba
\hat Q\hat \rho(t)=e^{\hat Q\hat L(t)t}\hat Q\hat \rho(t=0)+ \int ^{t}_{0}
dt'e^{\hat Q \hat L(t')t'}\hat Q \hat L(t')\hat P \hat \rho(t-t^{'})
\label{eq35}
\ea

If inserted into the first equation one obtains
\ba
\frac{d \hat P \hat \rho(t)}{dt}= \hat P \hat L(t)\hat P \hat \rho(t)+\hat P\hat L(t)
e^{\hat Q\hat L(t)t}\hat  Q\hat \rho(0)+\hat P \hat L(t)*
\notag\\
\int ^{t}_{0}dt'e^{\hat Q\hat L(t')t'}\hat Q \hat L(t')\hat P \hat \rho(t-t')
\label{eq36}
\ea  

This first order two-time integro-differential equation reduces to an ordinary one-time differential equation under one of the the following conditions:

\begin{itemize}

\item There is only one state $|\gamma\rangle$ in $E$ space. Then dim$\hat P=1$ and dim$\hat Q=0$.
As a consequence Eq.(33) reduces to
\ba
\frac{d \hat P \hat \rho(t)}{dt}=i\hat P[\hat P \hat \rho(t),\hat H]
\label{eq37}
\ea

\item The density operator at $t=0$ is such that $\hat P \hat \rho(0) \hat Q=0$, i.e. 
$\hat \rho(0)$ is block diagonal and furthermore $[\hat H_{E},\hat H_{SE}]=0$ in a basis of states in which $\hat H_{E}$ is diagonal. Then $\hat P \hat H \hat Q = 0$ and in the second terms of Eqs.(34a) and (34b), $ \hat P [\hat Q\hat  \rho(t),\hat H]=0$ and 
$\hat Q [\hat P \hat \rho(t),\hat H]=0$. This eliminates the second terms in Eqs.(34) which decouple.

\end{itemize}

Hence the evolution of the P-projected density operator $\hat P\hat \rho(t)$ is local in time and possesses the divisibility property. This result is again in agreement with the results obtained above and also with ref.~\cite{ck}. Finally the  evolution of the density operator in $S$  space $\hat \rho_{S}(t)$ is governed by 

\ba
Tr_{PE}\frac{d \hat P \hat \rho(t)}{dt}=iTr_{PE}\hat P[\hat P \hat \rho(t),\tilde H]
\notag
\ea
where $PE$ stands for the $P$ projection of $E$ space and $\tilde H=\hat H_{S}+\hat H_{SE}$.


\subsection{Memory effects and absence of divisibility: one-time approach}

A sufficient condition which induces the divisibility is obtained if $[\hat H_{E},\hat H_{SE}]=0$ in a basis of states in which $\hat H_{E}$ is diagonal. The violtion of divisibility is realized when $\hat H_{SE}$ possesses non-diagonal elements. Then the evolution of the density matrix is described by a master equation whose matrix elements for a fixed state $|\gamma\rangle$ in $E$ space depends on a unique time variable and takes the form 

\ba
\frac{d \hat \rho^{ik}_{S\gamma}(t)}{dt}=(-i)[\hat H_{d}^{\gamma},\hat \rho_{S\gamma}(t)]^{ik}+
(-i)\sum_{\beta \neq \gamma}[\Omega^{ik}_{\gamma \beta}(t)-\Omega^{ik}_{\beta \gamma}(t)]
\label{eq38}
\ea
where $\hat H_{d}^{\gamma}$ is the diagonal part in $E$ space of $\hat H$ for fixed $\gamma$ and  

\ba
\Omega^{ik}_{\gamma \beta}(t)=\sum_{j}\langle i \gamma|\hat H_{SE}|\beta j\rangle 
\langle j \beta |\hat \rho_{S}(t)|\gamma k\rangle
\notag\\
\Omega^{ik}_{\beta \gamma}(t)=\sum_{j}\langle i \gamma|\hat \rho_{S}(t)|\beta j\rangle
\langle j \beta |\hat H_{SE}|\gamma k\rangle
\label{eq39}
\ea

In the present formulation the master equation depends on a unique time variable although it describes a non divisible process. Physically it is the fact that the environment gets the opportunity to "jump" from a state $|\gamma\rangle$ to another state $|\beta\rangle$ which produces necessarily a time delay. This time delay induces the violation of the semi-group property when this delay is absent in the process. Here the strength of the violation is measured by the strength of the non-diagonal elements.\\ 

Finding physical systems which realize $[\hat H_{E},\hat H_{SE}]=0$ is certainly as difficult to realize as a rigorous Markovian quantum process.

\subsection{Entangled initial conditions}   

Consider the more general case for which initial correlations at $t_{0}$ are present~\cite{bu}. Then the initial density operator can be written as $\hat \rho (t_{0}) = |\Psi(t_{0})\rangle \langle \Psi(t_{0})|$ with$|\Psi(t_{0})\rangle= \sum_{i,\alpha} a_{i,\alpha}|i,\alpha\rangle$.
Using the same notations as above the component ($k_{1}, k_{2}$) of $\hat \rho^{k_{1} k_{2}}_{S}(t)$ reads 

\ba
\hat\rho_{S}^{k_{1} k_{2}}(t)= \sum_{i_{1},i_{2}}\sum_{\alpha_{1},\alpha_{2},\gamma} a_{i_{1},\alpha_{1}} a_{i_{2},\alpha_{2}}^{*}
U_{(i_{1}k_{1}),(\alpha_{1}\gamma)}(t,t_{0})U^{*}_{(i_{2}k_{2}),(\alpha_{2}\gamma)}(t,t_{0})|k_{1}\rangle \langle k_{2}|
\label{eq40}
\ea
In order to test the divisibility property of the system $S$ one introduces two arbitrary time intervals 
$[t,t_{s}]$, $[t_{s}, t_{0}]$ and looks  for an equality between the r.h.s. of Eq.(40) and the expression of the product of the evolution operators acting successively in the two intervals defined above. Equating both sides one gets 

\ba
\sum_{j_{1},j_{2}}\sum_{\eta}\sum_{\gamma}U_{(j_{1}k_{1}),(\eta\gamma)}(t,t_{s})U^{*}_{(j_{2}k_{2}),(\eta\gamma)}(t,t_{s})*
\notag\\
\sum_{\delta}U_{(i_{1}j_{1}),(\alpha_{1}\delta)}(t_{s},t_{0})U^{*}_{(i_{2}j_{2}),(\alpha_{2}\delta)}(t_{s},t_{0})|k_{1}\rangle \langle k_{2}|=
\notag\\
\sum_{\gamma} U_{(i_{1}k_{1}),(\alpha_{1}\gamma)}(t,t_{0})U^{*}_{(i_{2}k_{2}),(\alpha_{2}\gamma)}(t,t_{0})|k_{1}\rangle \langle k_{2}|
\label{eq41} 
\ea

A sufficient condition which leads to the formal equality of the two sides in Eq.(41) can be realized if the summation of the states are such that $|\delta\rangle=|\eta\rangle$. Then the summation over the intermediate states $j_{1},j_{2}$ on the l.h.s. of Eq.(41) can only be performed independently if the summation over $E$ space reduces to a unique state which allows the use of the closure property 
leading to the same summation over states in $E$ space on the right and left side of the equality:
the system $S+E$  stays in the same state over the whole interval of time $[t,t_{0}]$. 
The expression of the density matrix in $S$ space  in then given as 

\ba
\rho_{S}^{j_{1},j_{2}}(t)=\sum_{i_{1},i_{2}}a_{i_{1},\eta} a_{i_{2},\eta}^{*}
\langle j_{1}\eta|\hat U(t,t_{0})|i_{1} \eta\rangle 
\langle i_{2} \eta|\hat U^{*}(t,t_{0})|j_{2} \eta\rangle |j_{1}\rangle \langle j_{2}| 
\label{eq42} 
\ea
for a fixed state $|\eta\rangle$ in $E$ space. The Hilbert space of the total system $S+E$ has to reduce in practice to dimension $d+1$ where $d$ is the dimension of $S$. This is an exceptional situation and a sufficient condition for the realization of the property for correlated states at the initial time.
Problems related to the presence of initial correlations in an open quantum system coupled to an environment can lead to non positivity in the evolution of the system. This question has been investigated  in many works over the last years and comes out as a major difficulty, see f.i. ~\cite{rmma,schm} and refs. in there.

\subsection{Some further remarks concerning entanglement}  

It is the existence of the coupling $\hat H_{SE}$ between $S$ and $E$ which may generate entanglement between the system $S$ and the environment $E$. This coupling is also the source of time retardation (non-divisibility) effects in the time behaviour of the system $S$. 

A well known test concerning the time evolution of entanglement in an open quantum system has been proposed  as a conjecture by Kitaev and confirmed ~\cite{aco} which proves the so called "small incremental entangling" (SIE)~\cite{bra}. 

It was shown that in the absence of ancilla states the maximum time evolution of the von Neumann entropy 
$\Sigma_{S}(t)=-Tr \hat\rho_{S}(t)\log \hat\rho_{S}(t)$ verifies 

\ba
\Gamma_{max}=\frac{d\Sigma_{S}(t)}{dt}|_{t=0}\leq c \|\hat H\|\log \delta
\label{eq43} 
\ea
where $\delta=min(d_{S}, d_{E})$, the smallest dimension of $S$ and $E$ space, $\|\hat H\|$ is the norm of the interaction Hamiltonian $\hat H_{SE}$ and $c$ a constant of the order of unity.

In the case discussed above $\delta=1$, hence $\Gamma_{max}=0$ which shows that the entropy of the considered system does not change at the origin of time. There is no initial exchange of information between $S$ and $E$ space in this case. 

\subsection{Divisibility does not necessarily imply the commutation relation between the environment and the interaction Hamiltonians}

We discuss here the outcome of a simple model in order to show that the semi-group property can, under certain special conditions, also be valid for systems which do not satisfy the condition 
$[\hat H_{E},\hat H_{SE}]=0 $. Another example has been worked out elsewhere for non-Markovian systems~\cite{pol3}. 

Consider the case where the Hamiltonian $\hat H$ of the total system reads
  
\ba
\hat H= \hat H_{S}+\hat H_{E}+\hat H_{SE}
\label{eq44}
\ea

with 

\begin{center}
\ba
\hat H_{S}=\omega \hat J_{z} 
\notag \\
\hat H_{E}=\beta b^{+}b
\notag \\
\hat H_{SE}=\eta(b^{+}+b) \hat J^{2}
\label{eq45}
\ea 
\end{center}  
which corresponds to the case where this time $[\hat H_{S},\hat H_{SE}]=0 $, $b^{+},b$ are boson operators, $\omega$ is the rotation frequency of the system, $\beta$ the quantum of energy of the oscillator and 
$\eta$ the strength parameter in the coupling interaction between $S$ and $E$.

Since $\hat J_{z}$ and $\hat J^{2}$ commute in the basis of states $[|j m\rangle]$ the matrix elements of $\hat H$ in $S$ space read
  
\ba
\langle jm|\hat H|jm \rangle=\omega m+\beta b^{+}b+\eta j(j+1)(b^{+}+b) 
\label{eq46}
\ea 
  
The expression of the density operator $\hat \rho_{S}(t)$ at time $t$ is then obtained by taking the trace over the environment states of the total Hamiltonian $\hat \rho(t)$ leading to 

\ba
\hat \rho_{S}(t)=Tr_{E}\hat \rho(t) 
\label{eq47}
\ea 
whose matrix elements read  
  
\ba
\rho^{j m_{1}, j m_{2}}_{S}(t)=\rho^{j m_{1}, j m_{2}}_{0}(t)\Omega_{E}(j,j,t)
\label{eq48}
\ea 
with  
  
\ba
\rho^{j m_{1}, j m_{2}}_{0}(t)=e^{[-i\omega(m_{1}-m_{2})]t}/(2j+1)
\label{eq49} 
\ea  
The bosonic environment contribution can be put in the following form 
  
\ba
\Omega_{E}(j,j,t)=\sum_{n=0}^{n_{max}}\frac{1}{n!}\sum_{n',n^{"}}\frac{E_{n,n'}(j,t)
E^{*}_{n^{"},n}(j_{2},t)}{[(n'!)(n''!)]^{1/2}}
\label{eq50} 
\ea  
The results are exact. The Zassenhaus development formulated in Appendix D was used in order to work out the expressions ~\cite{za}. The expressions of the polynomials $E_{n,n'}(t)$ and $E^{*}_{n'',n}(t)$ are developed in Appendix E.\\

By simple inspection of the expressions in Appendix E it can be seen that the non-diagonal of $\rho^{j_{1} m_{1}, j_{2} m_{2}}_{S}(t)$ may cross zero when $t$ increases but oscillate and never reach and stay at zero whatever the length of the time interval which goes to infinity. 

More precisely the time dependence of $\hat \rho_{S}(t)$ is uniquely determined by the behaviour of oscillatory functions which depend on the  parameters $\omega$, $\gamma(j)$ and $\beta$. If these parameters induce commensurable oscillation periods one realizes that the density matrix can evolve periodically over periods which correspond to some $T$.

Hence if $\hat \rho_{S}(t)={\bf L_{(t,0)}}\hat \rho_{S}(0)$ then
\ba
\rho^{j m_{1}, j m_{2}}_{S}(t)={\bf L_{2T,T}}{\bf L_{T,0}}\rho^{j m_{1}, j m_{2}}_{S}(0)
\label{eq51}
\ea 
for $t=2T$. The expression reflects a semi-group property of $\rho^{j m_{1}, j m_{2}}_{S}(t)$ for selected time intervals $[nT,pT]$, $p \ge n$.
 
\section{Summary and conclusions}   
 
In the present work we examined different aspects of the evolution of open quantum systems. We first recalled the celebrated mathematical (classical) Markov process and examined different application of the concept in quantum physics. 

We then started from a phenomenological derivation of a master equation obeyed by the open system ~\cite{sh}. There it comes out that a Markovian behaviour can be approximatly realized under two conditions: a weak coupling between the system and its environment, a short memory correlation time in the environment compared to the characteristic evolution time of the system.

We recalled in a second step the recent formal derivation of a necessary and sufficient condition for the Markovian behaviour of a system which is tested at different times by means of stochastic processes ~\cite{pol1}. The result emphasizes the predictable impossibility of such systems to show a rigorous Markovian evolution. This does not come as a surprise and comforts the intuitive feeling that physical systems always react with a certain time delay to the action coming from  the outside.This time delay corresponds to a so called memory time, the system keeps track of the past, a process which takes a certain time and can be expressed in terms of trajectories. A process will be markovian iff it does not depend on the history of the evolution process ~\cite{pol1,sak}.   

In a second step we examined the case of systems which are not coupled to an environment through stochastic action and such that the Hamiltonians which govern the environment and the coupling between the environment and the system commute. In this case we presented sufficient conditions for which the evolution of a system possesses the semi-group (divisibility) property, a central characteristic of Markov processes. The derivation showed how non-Markovian effects manifest themselves in this case. If divisibility characterizes the system 
the environment follows the state in which the interaction started (i.e a fixed "trajectory", "channel"). If divisibility is not preserved as it is the  case in non-Markovian processes this property is no longer realized, i.e. the environment may choose different states in which it evolves leading to different histories. This fact explains the existence of a memory time, the time over which the system "feels" the  changes which happen in the environment. Recently an experiment led to the observation of "quantum jumps" between states in a driven quantum system ~\cite{min} which shows how different histories can be experimentally generated. Finally we addressed the problem which concerns the presence of entanglement at initial time between the states of the system of interest and the environment to which it is coupled and showed under which conditions the time evolution of the  system may remain divisible.   

In summary the Markovian or semi-group property of open quantum systems is generated through one of the following properties:

\begin{itemize}

\item the correlation time in the environment is small compared to the typical evolution time of the system.   

\item the environment follows a fixed "trajectory" in its evolution in time, whatever the interaction, stochastic or deterministic, weak or strong.

\item the divisibility property, i.e. the absence of time delay in the interaction between the system and its environment which may be due to specific properties of the environment and its interaction with the system.

\end{itemize}

In principle the density operators of systems which undergo memory effects are governed by two-time master equations. In the specific case we presented the dynamical equations show a one-time behaviour. The absence of time delays appears through the role played by non-diagonal matrix elements of the Hamiltonian which governs the interaction between the system  and its environment.This interaction can be arbitrarily large. Similarly to an example given in ref.~\cite{pol1} we worked out an academic example which confirms that divisibility is a property which does not necessarily imply Markovianity. 
The present work shows also that Markovianity and divisibility are properties which may rarely characterize the evolution of open quantum systems. 

\section{Appendix A: imposing the divisibility constraint}

Using the explicit expression of the super matrix $C$ given by Eqs.(23-24) the divisibility constraint in Eq.(26) for fixed states $(i_{1}, i_{2})$, $(k_{1}, k_{2})$ imposes the following relation 
 
\ba
\sum_{\alpha_{1},\alpha_{2},\gamma}d_{\alpha_{1},\alpha_{2}}U_{(i_{1}k_{1}),(\alpha_{1}\gamma)}(t-t_{0})
U^{*}_{(i_{2}k_{2}),(\alpha_{2}\gamma)}(t-t_{0})=
\sum_{j_{1},j_{2}}\sum_{\alpha_{1},\alpha_{2},\beta_{1},\beta_{2}}d_{\alpha_{1},\alpha_{2}} 
d_{\beta_{1},\beta_{2}}
\notag \\
\sum_{\gamma,\delta}U_{(j_{1}k_{1}),(\beta_{1}\delta)}(t-t_{s})U_{(i_{1}j_{1}),(\alpha_{1}\gamma)}(t_{s}-t_{0})
U^{*}_{(j_{2}k_{2}),(\beta_{2}\delta)}(t-t_{s})U^{*}_{(i_{2}j_{2}),(\alpha_{2}\gamma)}(t_{s}-t_{0}) 
\label{eq52}
\ea

In order to find a solution to this equality and without loss of generality we consider the case where the density matrix in $E$ space is diagonal. Then the equality reads 
 
\ba
\sum_{\alpha, \gamma}d_{\alpha,\alpha}U_{(i_{1}k_{1}),(\alpha \gamma)}(t-t_{0})
U^{*}_{(i_{2}k_{2}),(\alpha \gamma)}(t-t_{0})=
\sum_{j_{1},j_{2}}\sum_{\alpha,\beta}d_{\alpha,\alpha} d_{\beta,\beta}
\notag \\
\sum_{\gamma,\delta}U_{(j_{1}k_{1}),(\beta \delta)}(t-t_{s})U_{(i_{1}j_{1}),(\alpha \gamma)}(t_{s}-t_{0})
U^{*}_{(j_{2}k_{2}),(\beta \delta)}(t-t_{s})U^{*}_{(i_{2}j_{2}),(\alpha \gamma)}(t_{s}-t_{0}) 
\label{eq53}
\ea

A sufficient condition to realize the equality is obtained if $d_{\beta,\beta}=d_{\alpha,\alpha}$ and consequently if the weights $d$ on both sides are to be the same one ends up with $d_{\alpha,\alpha}=1$. This last condition imposes a unique state in $E$ space, say $|\eta \rangle$. In this case $d_{\eta,\eta}=1$ and Eq.(26) reduces to  
 
\ba
U_{(i_{1}k_{1}),(\eta \eta)}(t-t_{0})U^{*}_{(i_{2}k_{2}),(\eta \eta)}(t-t_{0})=
\sum_{j_{1}} U_{(i_{1}j_{1}),(\eta \eta)}(t_{s}-t_{0})U_{(j_{1}k_{1}),(\eta \eta)}(t-t_{s})
\notag\\
\sum_{j_{2}} U^{*}_{(j_{2}k_{2}),(\eta \eta)}(t-t_{s})U^{*}_{(i_{2}j_{2}),(\eta \eta)}(t_{s}-t_{0}) 
\label{eq54}
\ea
which proves the equality.

 \section{Appendix B: a special case of divisibility}  

Starting from the expression of the density operator given by Eqs.(21-24) we consider the case where 
$|c_{i}|=1/n$  for all $i$ where $n$ is the number of states in $S$ space and $d_{\alpha_{1},\alpha_{2}}=1/N \delta_{\alpha_{1},\alpha_{2}}$.

In this case the relation which imposes the divisibility constraint reads 
\ba
\frac{1}{Nn}\sum_{i\alpha,\gamma}U_{(ik_{1}),(\alpha\gamma)}(t,t_{0})U^{*}_{(ik_{2}),(\alpha\gamma)}(t,t_{0})=
\notag\\ 
\frac{1}{N^{2}n}\sum_{j_{1}j_{2}\beta,\delta}U_{(j_{1}k_{1}),(\beta\delta)}(t,t_{s})U^{*}_{(j_{2}k_{2}),(\beta\delta)}(t,t_{s})*
\notag\\
\sum_{i,\alpha,\gamma}U_{(ij_{1}),(\alpha\gamma)}(t_{s},t_{0})U^{*}_{(ij_{2}),(\alpha\gamma)}(t_{s},t_{0})
\label{eq55} 
\ea

The expression in the last line leads to  

\ba
\sum_{i,\alpha,\gamma}U_{(ij_{1}),(\alpha\gamma)}(t_{s},t_{0})U^{*}_{(ij_{2}),(\alpha\gamma)}(t_{s},t_{0})= N\delta_{j_{1},j_{2}} 
\label{eq56}
\ea 
 0
and finally the r.h.s. reduces to

\ba 
\sum_{j_{1}j_{2}\beta,\delta}U_{(j_{1}k_{1}),(\beta\delta)}(t,t_{s})U^{*}_{(j_{2}k_{2}),(\beta\delta)}(t,t_{s})=1/N\delta_{k_{1},k_{2}} 
\label{eq57}
\ea 
 
It is easy to observe that working out the l.h.s. of Eq.(57) leads to the same result.  

\section{Appendix C: different expressions of Eq.(29)} 

For a fixed unique state $\gamma$ the expressions of $\frac{d}{dt}\rho_{S1 \gamma}^{j_{1}j_{2}}(t)$ and $\frac{d}{dt}\rho_{S2 \gamma}^{j_{1}j_{2}}(t)$ given in Eq.(29) can be written as
 
\ba
\frac{d}{dt}\rho_{S1 \gamma}^{j_{1}j_{2}}(t)=(-i)\sum_{k_{1}k_{2}}
A^{j_{1}k_{1}}_{\gamma}\rho^{k_{1}k_{2}}_{S1 \gamma}(t)I^{k_{2}j_{2}} 
\label{eq58}
\ea
where $\hat I$ is the identity operator in S space and 
 
\ba
A^{j_{1}k_{1}}_{\gamma}=\langle j_{1}\gamma|\hat H_{S}|k_{1}\gamma\rangle + \langle j_{1}\gamma|\hat H_{E}+\hat H_{SE}|k_{1}\gamma\rangle
\label{eq59}
\ea 
and similar expressions for $\frac{d}{dt}\rho_{S2 \gamma}^{j_{1}j_{2}}(t)$. The matrix elements of $\hat H_{SE}$ in the second term on the r.h.s. of the expression of $A$ are generally non diagonal in $E$. They are diagonal if $\hat H_{E}$ and $\hat H_{SE}$ commute.

In symmetrized form the r.h.s. of the master equation reads 

\ba
(-i)\hat H \hat\rho_{S} +(i)\hat\rho_{S}\hat H=(\hat I-i\hat H)\hat\rho_{S}(\hat I+i\hat H)
-\hat\rho_{S}-\hat H \hat\rho_{S}\hat H
\label{eq60}
\ea 
 
\section{Appendix D: the Zassenhaus development} 
 
If $X=-i(t-t_{0})(\hat H_{S}+\hat H_{E})$ and $Y=-i(t-t_{0})\hat H_{SE}$

\ba
e^{X+Y}=e^{X}\otimes e^{Y}\otimes e^{-c_{2}(X,Y)/2!}\otimes e^{-c_{3}(X,Y)/3!}\otimes e^{-c_{4}(X,Y)/4!}...
\label{eq61}
\ea

where

\begin{center}
$c_{2}(X,Y)=[X,Y]$\\ 
$c_{3}(X,Y)=2[[X,Y],Y]+[[X,Y],X]$\\ 
$c_{4}(X,Y)=3[[[X,Y],Y],Y]+3[[[X,Y],X],Y]+[[[X,Y],X],X]$, etc.\\
\end{center} 
 
The series has an infinite number of term which can be generated iteratively in a straightforward way  
~\cite{ca}. If $[X,Y]=0$ the truncation at the  third term  leads to the factorisation of the $X$ and the $Y$ contribution. If $[X,Y]=c$ where $c$ is a c-number the expression corresponds to the well-known Baker-Campbell-Hausdorff formula.\\  

 \section{Appendix E: The bosonic content of the density operator}
 
The expressions of the bosonic contributions to the density matrix $\rho^{j m_{1}, j m_{2}}_{s}(t)$ are given by 
 
\ba
E_{n,n'}(j,t)=e^{-i\beta t}\sum_{n\geq n_{2},n_{3}\geq n_{2}}\sum_{n_{3}\geq n_{4},n'\geq n_{4}}(-i)^{n+n_{3}}
(-1)^{n'+n_{2}-n_{4}}
\notag\\
\frac{n!n'!(n_{3}!)^{2}[\alpha(t)^{n+n_{3}-2n_{2}}][\zeta(t)^{n_{3}+n'-2n_{4}}]}{(n-n_{2})!(n_{3}-n_{4})!
(n_{3}-n_{2})!(n'-n_{4})!}e^{\Psi(t)}
\label{eq62}      
\ea
and

\ba
E^{*}_{n^{"},n}(j;t)=e^{i\beta t}\sum_{n^{"}\geq n_{2},n_{3}\geq n_{2}}\sum_{n_{3}\geq n_{4},n\geq n_{4}}i^{n^{"}+n_{3}}
(-1)^{n+n_{2}-n_{4}}
\notag\\
\frac{n^{"}!n!(n_{3}!)^{2}[\alpha(t)^{n^{"}+n_{3}-2n_{2}}][\zeta(t)^{n+n_{3}-2n_{4}}]}{(n^{"}-n_{2})!(n_{3}-n_{2})!(n_{3}-n_{4})!(n-n_{4})!}e^{\Psi(t)}
\label{eq63}      
\ea

The different quantities which enter $E_{n,n'}(t)$ are 

\ba
\alpha(t)=\frac{\gamma(j)\sin\beta t}{\beta}
\label{eq64}      
\ea

\ba
\zeta(t)=\frac{\beta[1-\cos\gamma(j)t]}{\gamma(j)}
\label{eq65}      
\ea

\ba
\gamma(j)=\eta j(j+1)
\label{eq66}      
\ea

\ba
\Psi(t)=-\frac{1}{2}[\frac{\gamma^{2}(j)\sin^{2}(\beta t)}{\beta^{2}}+\frac{\beta^{2}(1-\cos\gamma(j)t)^{2}}{\gamma^{2}(j)}]                           
\label{eq67}      
\ea

\end{document}